\documentclass[iop]{emulateapj}
\usepackage{amsmath}
\usepackage{graphicx}
\usepackage{soul}

\begin{document}
\title{Simulated Analogs of Merging Galaxy Clusters Constrain the
  Viewing Angle}

\shorttitle{Analogs of Merging Galaxy Clusters}

\author{David Wittman}
\affiliation{Physics Department, University of California, Davis, CA 95616}
\author{B Hunter Cornell}
\affiliation{Physics Department, University of California, Davis, CA 95616}
\author{Jayke Nguyen}
\affiliation{Physics Department, University of California, Davis, CA 95616}
\affiliation{Physics Department, University of California, Berkeley,  CA 94720}
\email{dwittman@physics.ucdavis.edu}

\begin{abstract}
  A key uncertainty in interpreting observations of bimodal merging galaxy clusters is the unknown angle between the subcluster separation vector and the plane of the sky.  We present a new method for constraining this key parameter. We find analogs of observed systems in cosmological n-body simulations and quantify their likelihood of matching the observed projected separation and relative radial velocities between subclusters, as a function of viewing angle. We derive constraints on the viewing angle of many observed bimodal mergers including the Bullet Cluster (1E 0657-558) and El Gordo (ACT-CL J0102-4915). We also present more generic constraints as a function of projected separation and relative radial velocity, which can be used to assess additional clusters as information about them becomes available.  The constraints from these two observables alone are weak (typically $\gtrsim 70-75^\circ$ at 68\% confidence and $\gtrsim 55-60^\circ$ at 95\% confidence) but incorporate much more cosmological context than the classical timing argument, marginalizing over many realizations of
  substructure, peculiar velocities, and so on. Compared to the MCMAC code, which implements the timing argument on NFW halos, our constraints generally
  predict subcluster separation vectors closer to the plane of the sky. This is because in realistic mergers the subcluster velocity vectors are not entirely parallel to the separation vector (i.e, the mergers are not perfectly head-on). As a result,
 observation of a nonzero relative radial velocity does not exclude a separation vector in the plane of the sky, as it does in the head-on timing argument employed by MCMAC. 
  \end{abstract}


\section{Introduction}\label{sec-intro}

When galaxy clusters merge, they release a great deal ($\sim 10^{65}$
erg) of gravitational potential energy into other processes such as
galaxy motions, shocks, and particle acceleration. Merging clusters
thus provide fertile ground for studying a range of astrophysical
processes. The behavior of dark matter in such mergers is particularly
interesting because it may reveal clues to the particle nature of
dark matter. Momentum exchange between dark matter particles could
slow the dark matter halos relative to their galaxies, or could
scatter dark matter particles out of the halo entirely
\citep{Markevitch04,Kahlhoefer13,Robertson17Bullet,Robertson17aniso,Kim17}. 
If such a signal were to be found, a potential difficulty in translating
the observable (a positional offset or a reduction in mass-to-light
ratio) to a particle cross section lies in reconstructing the relative
speed of the subclusters at pericenter, and the time since pericenter,
from a single snapshot of the system taken hundreds of Myr later.

A standard way of estimating these quantities, called the timing
argument, originated in the 1950's \citep{Kahn1959}. The classic timing
argument assumes a head-on collision approximately in the plane of the sky, and
uses equations of motion appropriate to two point masses in an
expanding universe.  \citet{Dawson2012} presented a comprehensive update
to this approach, for example using Navarro-Frenk-White
\citep{NFW97} profiles rather than point masses, and
considering the full range of angles $\theta$ between the merger axis (which in this model defines
the separation vector as well as the velocity vectors) and the
plane of the sky.  Using this code, MCMAC,\footnote{\url{https://github.com/MCTwo/MCMAC}} he found that uncertainty in $\theta$ is a large,
previously unrecognized source of uncertainty in the other parameters.
This motivated subsequent work on ways to reduce uncertainty in
$\theta$, such as the inverse correlation between $\theta$ and the
polarization of synchrotron emission associated with the shocked
intracluster medium \citep{Ensslin1998,Ng2015}. The projected
separation between the shock and the center of mass also provides a
loose upper bound on $\theta$, as a sufficiently large $\theta$ would
imply a three-dimensional separation that would take longer than the
age of the universe for the shock to travel \citep{Ng2015}.

Here, we take an entirely different approach to constraining $\theta$,
based on n-body simulations of merging systems selected from a
cosmological volume. For any given observed system, the cosmological
simulation is searched for systems that could be analogs. Simulated
systems are ``viewed'' from all possible angles to obtain a likelihood
that the observed data match the simulated system, as a function of
$\theta$. This likelihood is then aggregated over all analogs to
obtain a final constraint on $\theta$.  The analog approach avoids some of the
biases of the timing argument by including processes that are not easily
captured analytically, such as dynamical friction (gas dynamics could also be included, but this first paper is based on simulations
without gas). Analogs also capture variations caused by the large-scale environment or by substructure within the two merging halos,
thus capturing uncertainties that are neglected in the timing argument. Notably, analogs naturally fold in a cosmologically motivated
range of impact parameters, which relates to bias (because the timing argument assumes the minimum possible impact parameter) as well as
uncertainty.

If analogs are an effective way to include more physics than is practical in a timing argument,
they can also serve as a bridge to more detailed merger simulations. Attempts to model observed
systems with "staged" hydrodynamic simulations typically use a few different impact parameters and a few
different initial velocities to span a range of possibilities. Analogs could provide
cosmologically motivated sets of parameters for this endeavor. Going further, analogs could provide
full initial conditions for detailed {\it ab initio} simulations of merging systems, which would thereby
include substructure as well.

The remainder of the paper is organized as follows.  In
\S\ref{sec-method} we outline our method. In \S\ref{sec-results} we
detail the resulting merger angle constraints, both for known systems
and more generic classes of systems. In \S\ref{sec-discussion} we discuss
the implications.

\section{Method}\label{sec-method}
In \S\ref{subsec-simulation} we give an overview of the simulation tools 
used to create our catalog of potential cluster analogs. In
\S\ref{subsec-cutoff} we explain the steps taken to find binary merging halo 
pairs within the simulation. In \S\ref{subsec-calculations} we explain the 
calculations used to quantify the theta dependence.

\subsection{Simulation}\label{subsec-simulation}
We used the publicly available Big Multidark Planck (BigMDPL) Simulation 
hosted on the \textit{cosmosim.org} website. This dark matter universe 
simulation has the largest box size in this collection of simulations, 
(2.5 Gpc/h)$^3$, which enables the highest number of analogs to be found 
within the simulation. The mass of each particle in the simulation is 
$2.359 \times 10^{10} M_{\odot}$, so even the least massive clusters we consider 
in our catalog have over 250 particles. The simulation starts at redshift $z = 100$, 
running until the present day ($z = 0$) \citep{BigMDPL2013}. 

The galaxy clusters in the BigMDPL simulation were created using the Rockstar Halo finding algorithm \citep{Rockstar2013}.  The availability of multiple time steps in this simulation allows us to filter the data to exclude halo pairs based on whether they are infalling for the first time, or have already passed through pericenter, as described below. In this paper we model observed systems whose x-ray morphology indicates that they are post-pericenter, but this method would just as well support study of pre-pericenter systems by taking the complement of the post-pericenter set of analogs.

\subsection{Data Cuts}\label{subsec-cutoff}
To create our catalog of merged halo pairs, we first filter for halos with $ M > 6 \times 10^{13}\,M_{\odot}$ in all time steps. This is an appropriate mass range for the observed subclusters we seek to match, which are several times $10^{14}\,M_{\odot}$ and up, after accounting for some uncertainty on the observed masses. Within this mass-selected subset, we found all halo pairs with separations of $\le 5$ Mpc at each cluster's respective redshift to get a list of all possible merger analogs for each cluster. This 5 Mpc cut was generous; after imposing the further cuts described below we found no pair separated by more than 1.9 Mpc. 

After finding all the possible analog pairs, we remove any non binary systems. This is done by removing any pairs with a halo that is in our list of potential mergers more than once. We do this in order to restrict our analysis to pairs that are largely binary, since our analysis is limited to real mergers that are approximately binary.

We then filter for halo pairs that are post-pericenter and have a small pericenter distance ($r_p$) because the physical systems we are modeling exhibit x-ray morphology and radio relics that indicate this must be the case. \citep{GolovichRelicSampleData}. Specifically, we find halo pairs that passed within $r_p < 0.5$ Mpc at any point in time. 
The $r_p < 0.5$ Mpc cut is a conservative initial upper bound on the pericenter distance of the observed systems considered here, which will be further reduced after presenting further evidence in \S\ref{subsec-generic}. We stress the distinction between pericenter distance and impact parameter, which is defined at very early times as the component of the separation vector that is perpendicular to the velocity vector and which is commonly quoted in simulation papers because it is an initial condition of the simulation.  Because of gravitational attraction, the pericenter distance can be much ($>3$ times) smaller than the impact parameter \citep{Zhang2016PericenterImpactParameter}. The impact parameters of many of the clusters considered in this paper have been estimated by hydrodynamic simulations to be on the order of hundreds of kpc \citep{Molnar2015Gordo,Molnar2017,Zhang2015ElGordoHydro,Lage2014}. This implies that $r_p$ is several times smaller, well within this initial cut of 0.5 Mpc. We discuss the $r_p$ cut further in \S\ref{subsec-generic}.

We also remove any pairs that have merged multiple times in their history by filtering out any halos that were within $0.5$ Mpc of each other two or more times. This tends to be a fairly rare occurrence, as the period T (time between pericenters) is large for most clusters \citep{Dawson2012,Kim17,Ng2015}.
Pairs are determined to be post-pericenter if their smallest recorded separation is in a previous time-step, or their smallest separation is in the current time-step but they also have an outbound velocity. This latter set of pairs corresponds to recent mergers whose pericenter fell between the current time-step and the previous one. Pericenter distance is interpolated between time-steps by taking the component of the separation vector perpendicular to the velocity vector, {\it at the closest recorded separation}. After collecting all pairs that survive these cuts, we perform the calculations detailed below to constrain $\theta$ for any given observed system.

\begin{table}
\begin{tabular}{ccc}
Symbol & Description\\
\hline
$M_1$ & Virial Mass for Subcluster 1\\
$M_2$ & Virial Mass for Subcluster 2\\
$\Delta v_r$ & relative radial velocity of the two subclusters\\
$d_{\rm proj}$ & projected separation between subclusters\\

\end{tabular}
\caption{Observables used}
\label{tab-observables}
\end{table}

\subsection{Probability Calculations}\label{subsec-calculations}
\begin{figure}
\centerline{\includegraphics[scale=.4]{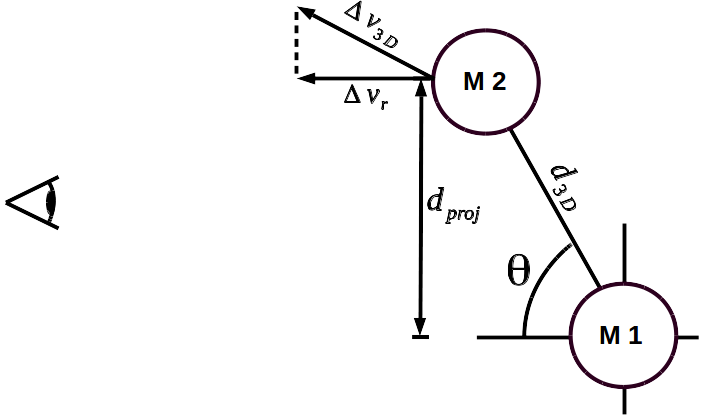}}
\caption{Definition of $\theta$, $\Delta v_r$, and $d_{\rm proj}$ for observed systems. The azimuthal angle $\phi$ (not shown) corresponds to rotating the observer out of the page, about the $M_1-M_2$ separation vector.} 
\label{fig-LOSDrawing}
\end{figure}

For a specific observed system we have a data vector $D$ consisting of the elements listed in Table~\ref{tab-observables}, along with their associated uncertainties. To calculate the likelihood that each analog $A$ fits the observed system, we define a spherical coordinate system such that a hypothetical line of sight lies $\theta$ radians from the sub-cluster separation vector, with $\phi$ encoding the azimuth about that vector. Our definition of $\theta$ is represented graphically in Figure \ref{fig-LOSDrawing}. The origin of $\phi$ was arbitrarily chosen to be either sub-cluster, but is not physically meaningful and is chosen based on computational convenience. For a potential analog system $A$ in the simulation, we compute the probability of $D$ given $A$ and a hypothetical line of sight along $\theta$ and $\phi$:
\begin{equation}
P(D|A,\theta,\phi)\propto P(M|A)P(d_{\rm proj}|A,\theta)P(\Delta v_r|A,\theta,\phi).
\end{equation}
where each term is explained in further detail below. Note that we use the proportionality symbol throughout because the probabilities will be renormalized at the end.

{\it Mass.} We first quantify how well the analog system matches the observed subcluster masses $M_1$ and $M_2$. This is the fastest step because it requires no $\theta$ or $\phi$ grid, and it could potentially be used to avoid spending further CPU time on analogs that are certain to be poor matches at all $\theta$ and $\phi$. However, we found that this potential speedup was not necessary for this particular halo catalog.

There are two distinct possibilities for matching the mass of the observed system: analog subcluster $a$ can match $M_1$ with analog subcluster $b$ matching $M_2$, or vice versa. We therefore calculate 
\begin{equation}
P(M|A) \propto P(M_1,M_2|A) + P(M_2,M_1|A)
\end{equation}
with
\begin{multline}
P(M_1,M_2|A) \propto \\ \exp{ -\frac{(M_1-M_{a})^2}{2\sigma_{M_1}^2}} \exp{ -\frac{(M_2-M_{b})^2}{2\sigma_{M_2}^2}} 
\end{multline}
and 
\begin{multline}
P(M_2,M_1|A) \propto \\ \exp{ -\frac{(M_2-M_{a})^2}{2\sigma_{M_2}^2}} \exp{ -\frac{(M_1-M_{b})^2}{2\sigma_{M_1}^2}} .
\end{multline}

{\it Projected separation.} For a given analog system, we find the length of the separation vector, $d_{3d}$. Taking $d_{\rm proj}$ as the observed central value of the projected separation and $\sigma_{d_{\rm proj}}$ as its uncertainty, we compute
\begin{equation}
P(d_{\rm proj}|A,\theta) \propto \exp{ -\frac{(d_{\rm proj}-d_{3d}\sin\theta)^2}{2\sigma_{d_{\rm proj}}^2}} 
\end{equation}
over a uniform grid in $\cos\theta$ to reflect a set of observers placed randomly about the sphere. Note that constraints on $d_{\rm proj}$ in the observational literature are heterogeneous. Some papers quote the distance between brightest cluster galaxies (BCGs) which has a small nominal uncertainty thus yielding fewer likely analogs, while others quote the less certain lensing centroid separation which yields more analogs. The lensing separation is more directly comparable to the halo separation we pull from BigMDPL.

{\it Relative velocity.} At each $\theta$ we sample a uniform grid in $\phi$ and project the relative velocity vector onto the line of sight to obtain $\Delta v_{r,A}$. If published measurements are described in terms of a central value $\Delta v_r$ and symmetric error bar $\sigma_{\Delta v_r}$ we assume a Gaussian distribution, 
\begin{equation}
P(\Delta v_r|A,\theta,\phi)\propto \exp{ -\frac{(\Delta v_r-\Delta v_{r,A})^2}{2\sigma_{\Delta v_r}^2}}.
\end{equation}
Note that this may be a strong function of $\phi$ because the subclusters are not necessarily on radial orbits. Where the published upper and lower limits are markedly asymmetric about the central value, we use a Gaussian with error bars corresponding to the larger of the upper and lower error bars. We then marginalize over $\phi$ to obtain $P(\Delta v_r|A,\theta)$. If desired, one could marginalize over $\theta$ as well to obtain an overall likelihood of the simulated system matching the observed system, but we are more interested in the behavior as a function of $\theta$.

{\it Assembling the results.} The previous steps determine the likelihood of the data matching the $i$th simulated analog, $A^i$ (we use superscripts to avoid confusion with labeling subclusters within an analog). We then sum over potential analogs in the cosmological volume: $P(D|\theta)\propto \sum_{i} P(D|A^i,\theta)$ and normalize to unit probability.  This process is repeated independently
for each observed system. 

\begin{figure}
\centerline{\includegraphics[scale=.7]{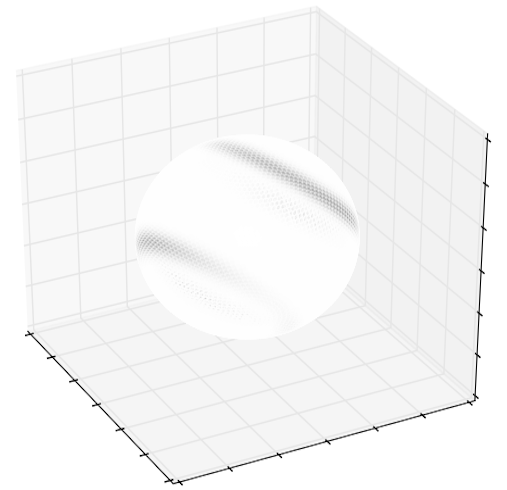}}
\caption{Lines of sight are placed randomly around a sphere, looking ``in'' at a simulated merger, and are 
shaded according to the likelihood of matching a particular observed system's projected separation and relative radial velocity. Using the subcluster separation vector (not shown) to define a colatitude $\theta$, matches to the projected separation can occur only along two bands of colatitude. Likelihood varies with longitide along those bands because the subcluster velocities are not parallel to the separation vector. The Cartesian grid shows the BigMDPL coordinates.} 
\label{fig-3D_Sphere}
\end{figure}
 
\section{Results}\label{sec-results}

In \S\ref{subsec-generic} we scan through the parameter space of
observables and show the viewing angle constraints at different points in
this space. This gives a general sense of how the constraints vary, and 
allows readers to look up constraints for systems not specifically listed 
in this paper. In \S\ref{subsec-individual} we present viewing 
angle constraints for specific systems for which the relevant input data 
can be found in the literature. In \S\ref{subsec-constraintcomparison} 
we compare those constraints with previously 
published viewing angle estimates of those clusters.

\subsection{Viewing angle constraints for generic merging clusters} \label{subsec-generic}

\begin{figure*}
\centerline{\includegraphics[scale=0.4]{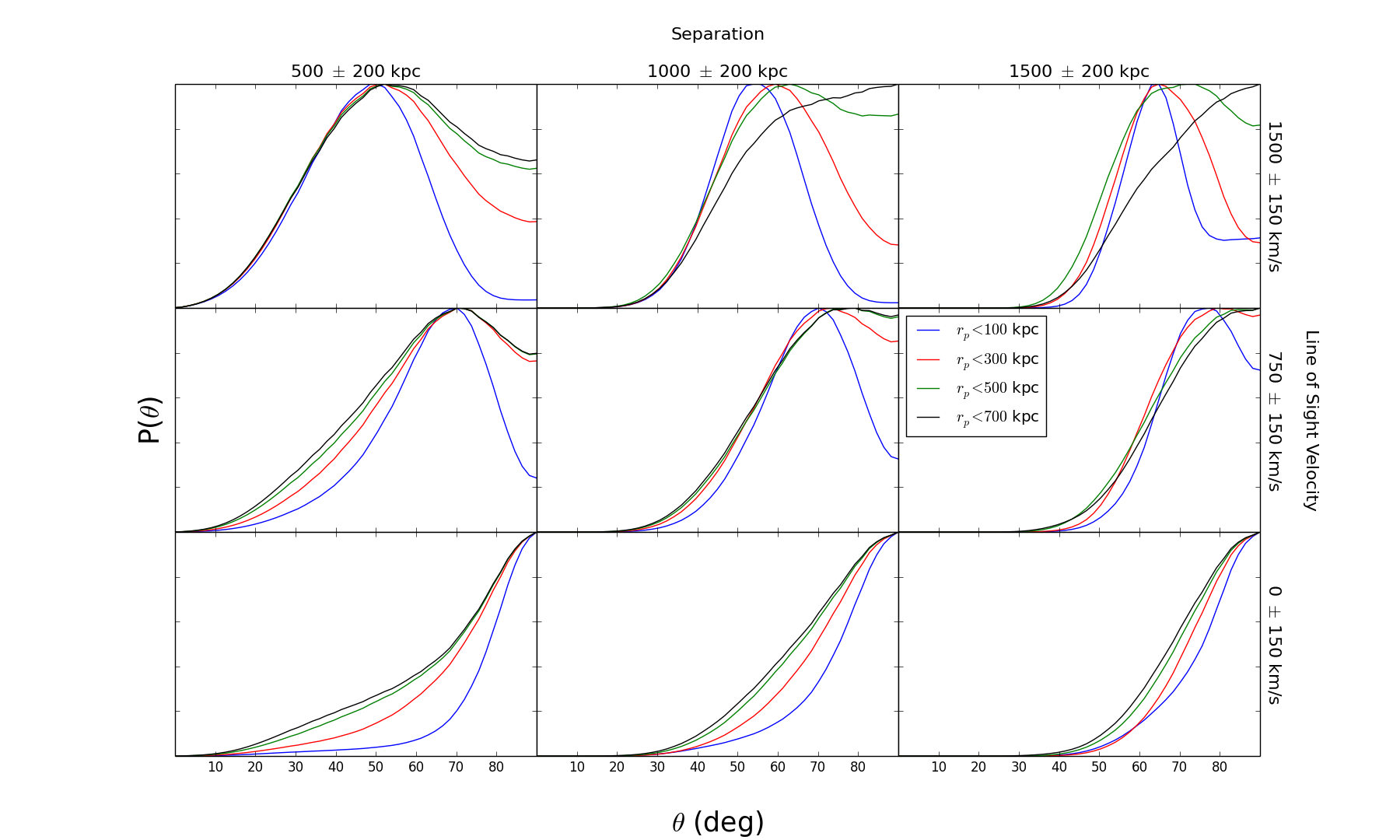}}
\caption{Viewing angle constraints for a range of $\Delta v_r$ and $d_{\rm proj}$ spanning $r_p$ space. In each panel the vertical axis is the likelihood normalized to unit maximum.} 
\label{fig-GeneralImpactParameter}\end{figure*}

We begin by presenting generalized merger scenarios at $z=0$. Figure~\ref{fig-GeneralImpactParameter} shows three illustrative values of $d_{\rm proj}$, each with three illustrative values of $\Delta v_{r}$; focus on the black curves first. Reading from the bottom row of panels upward, we find the expected trend in which larger $\Delta v_{r}$ implies a merger axis more along the line of sight. Reading from the left column of panels rightward, we find the expected trend in which larger $d_{\rm proj}$ implies a merger axis more in the plane of the sky. Finally, by comparing different colors within a panel we see the effect of different cuts on pericenter distance when forming the halo pair catalog. We see that the viewing angle constraint is tighter when a stricter cut is imposed. This makes sense, because a stricter cut provides a more homogeneous sample of nearly head-on mergers. Note that all colors are normalized to the same maximum value, but the more generous cuts do include more analogs. 

The most dramatic dependence on pericenter distance occurs in the upper right panel, where larger $r_p$ shifts the peak location in addition to broadening the distribution. We are not aware of any observed mergers that have such large projected separation {\it and} such large relative velocity, but we include this panel for completeness. Head-on mergers (those with small $r_p$, blue curve) are unlikely to appear in this panel because of the tradeoff between projected separation and radial velocity; if they appear, it is at a specific viewing angle where we can see most of the 3-d separation {\it and} most of the 3-d velocity. In contrast, orbits that are more circular (large $r_p$, black curve) will easily match such hypothetical observations, especially so when the separation vector is close to the plane of the sky.

For observed systems we must choose an appropriate set of analogs despite having limited knowledge of the pericenter distance. This paper applies the technique to  observed systems with highly disturbed X-ray morphologies, which rules out large pericenter distance.  For example, hydrodynamical modeling of ZwCl 0008.8+5215 yields an estimated impact parameter of $400\pm100$ kpc \citep{Molnar2017}; of CIZA J2242.8+5301, 120 kpc \citep{Molnar2017CIZA}; and of the Bullet Cluster, $256\pm 35$ kpc \citep{Lage2014}. For El Gordo, \citet{Molnar2015Gordo} find 300 kpc while \citet{Zhang2015ElGordoHydro} and \citet{Zhang2018ElGordoHydro} present both 300 kpc and 800 kpc scenarios. With the impact parameter generally exceeding the pericenter distance by a factor of several, $r_p>300$ kpc is ruled out in all these cases. One could argue that, with the exception of one group's studies of one system, even $r_p>100$ kpc is ruled out for these observed mergers. Nevertheless, as one of our goals is to avoid a bias toward assuming head-on mergers, we choose the more conservative cut, $r_p<300$ kpc, for the remainder of this paper. Figure~\ref{fig-GeneralImpactParameter} demonstrates that this choice does not shift the most likely viewing angle, though of course it broadens the distribution.

\begin{figure*}
\centerline{\includegraphics[scale=0.4]{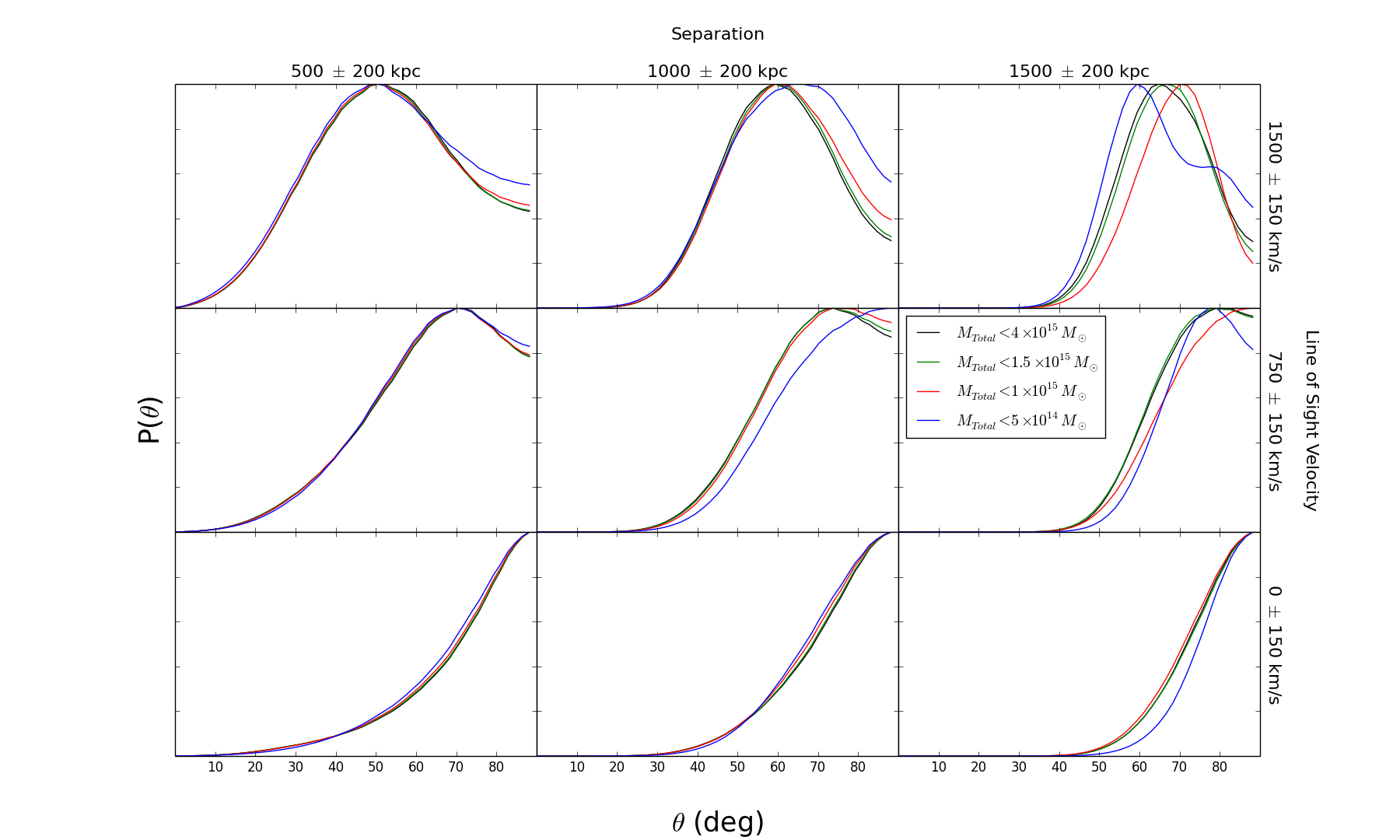}}

\caption{Viewing angle constraints for a range of $\Delta v_r$ and $d_{\rm proj}$ spanning total cluster mass space. In each panel the vertical axis is the likelihood normalized to unit maximum.} 
\label{fig-GeneralMass}\end{figure*}

Figure~\ref{fig-GeneralMass} breaks the analogs into subcategories by total mass, while including all analogs with pericenter distance up to 300 kpc. The mass dependence is generally weak, with many of the curves being nearly indistinguishable. A stronger dependence appears in the hypothetical case of large projected separation {\it and} large relative velocity, but we are aware of no observed system with these extreme properties. Therefore, observed systems can still be well constrained by marginalizing over all masses. This is useful because reliable mass estimates are difficult for merging clusters, with X-ray and dynamical masses biased high. Note that Figure~\ref{fig-GeneralImpactParameter} in fact marginalized over simulated clusters of all masses; Figure~\ref{fig-GeneralMass} provides justification for doing so. Finally, note that in the upper right panel of Figure~\ref{fig-GeneralMass}, the blue curve exhibits more complicated behavior than all other curves. This may be due to the paucity of analogs: low-mass systems are likely to exhibit high speeds only at smaller separations.

\subsection{Individual clusters}\label{subsec-individual}

Our primary source for data on merging clusters is the Merging Cluster Collaboration (MCC), which has published an imaging and spectroscopic survey of radio-selected mergers \citep{GolovichRelicSampleData}. For the mergers in their sample that can be plausibly approximated as binary, we draw $\Delta v_r$ and $d_{\rm proj}$ measurements from that paper. We supplement this with a few additional well-known merging systems. All quantities used as inputs for our observed system parameters are listed in Table~\ref{tab-clusterinputs}.

\begin{figure*}
\centerline{\includegraphics[scale=0.48]{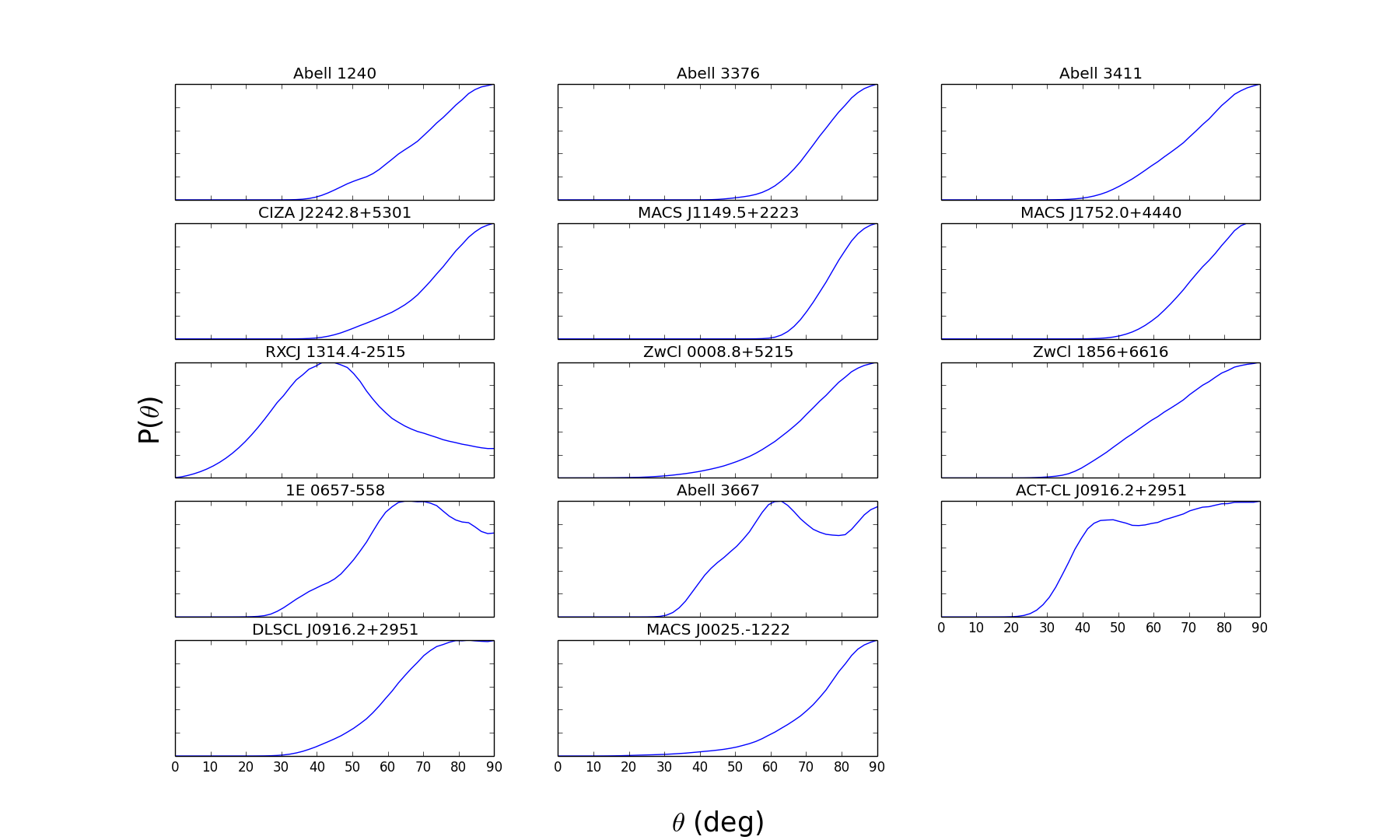}}
\caption{Likelihood of $\theta$ for each cluster. In general, the most likely viewing angle is at $\theta = 90^\circ$ unless the cluster has a substantial $\Delta v_r$ and small $d_{\rm proj}$. The multi-modality of Abell 3667 is discussed in \S\ref{subsec-individual} and Figure \ref{fig-revised_inputs}.}
\label{fig-thetaconstraints}
\end{figure*}

Figure~\ref{fig-thetaconstraints} shows the likelihood for $\theta$ for each of the 14 systems listed in Table~\ref{tab-results}. \S\ref{sec-radioselected} briefly summarizes the data sources and results for each MCC system, and \S\ref{sec-otherselected} does so for the remaining systems.
Unless otherwise noted, $\Delta v_r$ and $d_{\rm proj}$ measurements are drawn from \citet{GolovichRelicSampleData}. Mass constraints are drawn from lensing measurements in the literature where available. Where not available, we estimate dynamical masses from the galaxy member velocities published in \citet{GolovichRelicSampleData}. Although dynamical masses are likely to be biased high during a merger \citep[by a factor of a few;][]{Pinkney96,Takizawa10}, the dependence on mass is relatively weak (\S\ref{subsec-generic}). Dynamical masses are therefore sufficient for this procedure.

\begin{table*}
\centering
\begin{tabular}{lcccc}
Cluster & $M_1$ \& $M_2(10^{14} M_\odot)$ & $d_{\rm proj} (Mpc)$ & $\Delta v_r (km/s)$ \\
\hline

Merging Cluster Collaboration \citep{GolovichRelicSampleData}\\
\hline
Abell 1240 & 4.19 (.99)  4.58 (1.40) & 0.987 (.099) & 394 (118)\\
Abell 3376 & 3.0 (1.7) 0.9 (0.8) & 1.096 (0.066) & 181 (147)\\
Abell 3411 & 18.0 (5.0) 14.0 (5.0) & 1.4 (.20) & 80 (170)\\
CIZA J2242 & 11.0 (3.7) 9.8 (3.8) & 1.3 (.13) & 69 (190)\\
MACSJ1149 & 16.7 (1.3) 10.9 (3.4) & 0.995 (0.064) & 302 (219)\\
MACSJ1752 & 13.22 (3.14) 12.04 (2.59) & 1.145 (0.115) & 114 (158)\\
RXCJ1314 & 6.07 (1.8) 7.17 (2.8) & 0.54 (0.23) & 1523 (149) \\
ZWCI0008 & 5.7 (2.8) 1.2 (1.4) & 0.924 (0.243) & 92 (164) \\
ZWCI1856 & 9.66 (4.06) 7.6 (4.05) & 0.925 (0.093) & 189 (267) \\
\hline
Other Clusters\\
\hline

1E 0657 & 15.0 (1.5) 1.5 (0.15) & 0.72 (.075) & 616 (250)\\
Abell 3667 & 14.59 (2.74) 13.26 (2.67) & 1.0 (0.1) & 525 (108)\\
ACT-CL J0102 & 13.8 (2.2) 7.8 (2.0) & 0.7 (0.1) & 476 (242) \\
DLSCL J0916 & 3.1 (1.2) 1.7 (2.0) & 1.0 (.14) & 670 (330)\\
MACSJ0025 & 2.5 (1.7) 2.6 (1.4) & 0.518 (0.116) & 100 (80)\\

\end{tabular}
\caption{Values of observables used for each cluster \\ 
{\scriptsize Values in parentheses are the the symmetric error bars for each constraint used in the probability calculations}}
\label{tab-clusterinputs}
\end{table*}

\subsubsection{MCC Clusters}\label{sec-radioselected}
In general, the Merging Cluster Collaboration clusters, which are radio-selected, exhibit a suppression of $P(D|\theta)$ at low angles. This is driven by the large and highly significant values of $d_{\rm proj}$, which eliminate analog lines of sight close to the separation vector. The most likely viewing angle tends to place the separation vector in the plane of the sky ($\theta=90$). Of course, spherical geometry alone dictates that a randomly placed separation vector is more likely to be perpendicular to the line of sight than along the line of sight; however if this were the only effect we would see a weak constraint, $\theta>18^\circ$ at 95\% confidence, rather than the $\theta\gtrsim 50^\circ$ typical of the MCC systems in Table~\ref{tab-results}. We conclude that radio selection favors mergers in the plane of the sky, as one would expect from the low surface brightness of face-on radio emission in magnetohydrodynamical simulations of merging clusters \citep{Skillman13}. This bias is desirable for many merging-cluster studies because it enables measurement of separations between gas, galaxies, and dark matter.

RXCJ 1314.4-2515 is the only system where the viewing angle constraint departs from this pattern, as one can see by scanning the top three rows of Figure~\ref{fig-thetaconstraints}. We now offer brief comments on each system, primarily to note the sources of the measurements, with extra commentary on RXCJ 1314.4-2515.

{\it Abell 1240}. We obtain all the data for this cluster from \citet{GolovichRelicSampleData}. 

{\it Abell 3376}. We obtain all data for this cluster from \citet{Monteiro2017Abell3376}. We note that BigMDPL may contain additional analogs of this system, because the lower mass limit we imposed when harvesting halo pairs from BigMDPL is only about one standard deviation below the observed mass of this system's least massive subcluster. We do not expect this to affect the results because the viewing angle constraints depend only weakly on mass (\S\ref{subsec-generic}).

{\it Abell 3411}. Data for this cluster are from \citet{vanWeeran2017Abell3411}. 

{\it CIZA J2242.8+5301}. We use the lensing mass estimates from \citet{JeeCIZA2014}, with velocity and transverse separation data from \citet{DawsonCIZA2014}. 

{\it MACS J1149.5+2223}. We obtain all data for this cluster from \citet{GolovichMACS1149}, who discovered a second massive subcluster in addition to the well studied cluster in the Frontier Field.   

{\it MACS J1752.0+4440}. We obtain all the data for this cluster from \citet{GolovichRelicSampleData}. 

{\it RXCJ 1314.4-2515}. This is an instructive test case, because it is the only bimodal cluster in the \citet{GolovichRelicSampleData} sample for which $\Delta v_r$ is significantly nonzero ($1523~\pm$ 149 km/s). This is evident not from the overall histogram presented in \citet{GolovichRelicSampleData}, but from a spatially resolved velocity map (Golovich et al, in prep). Given a significant line-of-sight velocity {\it and} a nontrival projected separation, one would expect an intermediate viewing angle that allows us to see much of the true velocity as well as much of the true separation, and indeed the likelihood peaks near $\theta = 42^\circ$, quite unlike the other systems in this sample. Put another way, this system belongs in the upper left panel of Figure~\ref{fig-GeneralImpactParameter} while the other systems in this sample belong in the lower middle panel of that figure.

Despite the high line-of-sight velocity, $\theta=90^\circ$ is not ruled out; this is because analog systems can have relative velocities nearly perpendicular to their separation vectors. At the same time, $\theta=0 ^\circ$ (separation vector strictly along the line of sight) {\it is} ruled out because there is not enough uncertainty in the $d_{\rm proj}$ measurement to allow $d_{\rm proj}=0$ kpc. 

{\it ZwCl 0008.8+5215}. We draw all data for this cluster from \citet{GolovichZwCl0008}, who describe it as an older and less massive Bullet cluster. 

{\it ZwCl 1856+6616}. We obtain all the data for this cluster from \citet{GolovichRelicSampleData}.

\subsubsection{Other Merging Clusters}\label{sec-otherselected}

{\it 1E 0657-558 (Bullet Cluster)}. To facilitate later comparison with MCMAC results we use the inputs listed by \citet{Dawson2012}, who in turn drew from \citet{BarrenaBullet2002}, \citet{BradacBulletLensing2006}, and \citet{Springel2007}. Note that this literature contains measurements of $d_{\rm proj}$ based on the BCG separation as well as on the separation between lensing centers. The value of $d_{\rm proj}$ is similar in either case, but the uncertainty is larger for lensing. For Figure~\ref{fig-thetaconstraints} we use the lensing value and uncertainty, as does \citet{Dawson2012}. 

Remarkably, the likelihood does not peak at $\theta=90^\circ$. This is unlike the MCC radio-selected clusters (with the exception of RXCJ 1314.4-2515) and is driven by the substantial $\Delta v_r$ of 615 km/s. The likelihood peaks at $72^\circ$, or $18^\circ$ from the plane of the sky. This is consistent with the detailed hydrodynamical simulations of \citet{LageFarrar2014}, who find that at the time of best fit, the separation vector is $22^\circ$ from the plane of the sky (Lage, private communication). Our uncertainty on $\theta$ is large enough, however, to allow a wide range of scenarios. 

\begin{figure*}
\centerline{\includegraphics[scale=0.48]{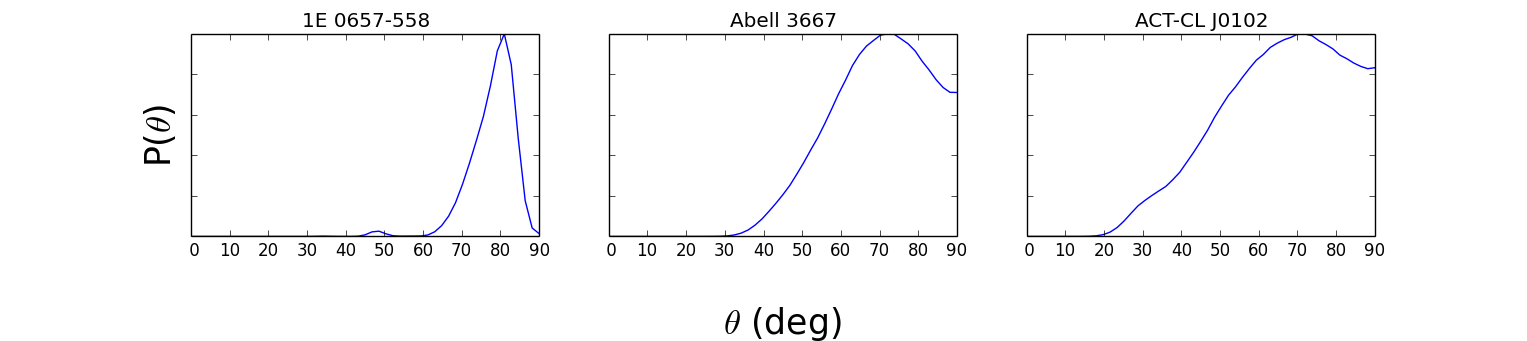}}
\caption{Likelihoods for the Bullet Cluster, Abell 3667, and El Gordo with revised parameter inputs. For the Bullet Cluster, we use the BCG separation, rather than the more uncertain lensing separation, for $d_{\rm proj}$ \citet{BradacBulletLensing2006}. For Abell 3667, the mass matching condition has been eliminated to provide more potential analogs. For El Gordo, we use analogs from a redshift of $z = 0$ in order to provide more potential analogs. These last two techniques provide smoother curves while maintaining the overall shape.}
\label{fig-revised_inputs}
\end{figure*}

Some wiggles are evident in the likelihood for this cluster. Because only a handful of analogs match the observations well, each analog can cause a noticeable bump. To illustrate this further, we rerun the algorithm with the 25 kpc uncertainty in the BCG $d_{\rm proj}$ rather than the 75 kpc uncertainty in the lensing $d_{\rm proj}$. The resulting likelihood, plotted in the left panel of Figure~\ref{fig-revised_inputs}, is now quite smooth but narrowly peaked about $\theta=80^\circ$. This is because a single analog system now contributes most of the weight, and we are essentially seeing $P(\theta)$ for that one analog. This illustrates one of the tradeoffs of the analog method: the number of usable analogs is inversely related to the measurement precision. This limitation is not seen with the timing argument, which can generate a continuous family of models. However, one can always obtain more analogs by going to larger simulation volumes. 

We note that Bullet analogs are rare in BigMDPL not because of the high mass, but because $\Delta v_r$ is constrained to such a narrow range, $\pm 80$ km/s, by \citet{BarrenaBullet2002}. (This is unusually precise among the systems considered here, especially given the number of galaxy spectra in the Bullet subcluster.) To test this, we reran the algorithm with a larger $\Delta v_r$ uncertainty (while maintaining the large lensing uncertainty on $d_{\rm proj}$), and the likelihood is indeed smoother than in Figure~\ref{fig-thetaconstraints} but with the same overall character. Because BCG position measurements are quite precise, using the
BCG separation for $d_{\rm proj}$ narrows the analog pool just as a precise $\Delta v_r$ measurement does. However, as BigMDPL does not contain galaxies, we do not recommend using the BCG position. Although BCG position may be a proxy for halo position, one would need to know the scatter in that relationship to compute likelihoods for analogs in BigMDPL. If BCG positions were to be used, it may be better to apply them directly to a simulation with galaxies.

{\it Abell 3667}. Both subclusters are ostensibly quite massive because the only available mass estimate is a dynamical mass from \citet{Owers2009Abell3667}, who also list the relative velocity and separation distances of the subclusters along with their associated uncertainties. This creates a situation where the likelihood is multimodal because only a few simulated systems match well. We produce a revised estimate by marginalizing over all masses, and plot the result in the middle panel of Figure~\ref{fig-revised_inputs}. The additional analogs help smooth the likelihood. After smoothing, the likelihood is remarkably similar to that of the Bullet cluster with $d_{\rm proj}$ from lensing, peaking at $\theta=72^\circ$. 

{\it ACT-CL J0102-4915 (El Gordo)}. The relative velocity and uncertainty are taken from the \citet{Ng2015} reanalysis of data from \cite{Sifon13}. Lensing mass estimates are taken from \citet{JeeGordo}. There are few massive clusters in BigMDPL at this high redshift, $z = 0.87$, so just a few analogs dominate and the curve is again multimodal. In order to smooth the curve, we rerun the analysis at $z = 0$ where there are many more potential analogs. The resulting curve is shown in Figure~\ref{fig-revised_inputs}. It appears similar to those of the Bullet cluster (with $d_{\rm proj}$ from lensing) and Abell 3667 (after smoothing).

{\it DLSCL J0916.2+2951 (Musketball Cluster)}.  Lensing masses, projected separation distance, and relative velocites are taken from \citet{Dawson11}. The viewing angle likelihood is more like the MCC radio-selected systems than the previous three systems.

{\it MACSJ0025.4-1222}. We obtain all data for this cluster from \citet{bradac2008}. The viewing angle likelihood is similar to that of DLSCL J0916.2+2951 and the MCC radio-selected systems.

\subsubsection{Summary of individual clusters}

\begin{figure*}
\centerline{\includegraphics[scale=0.48]{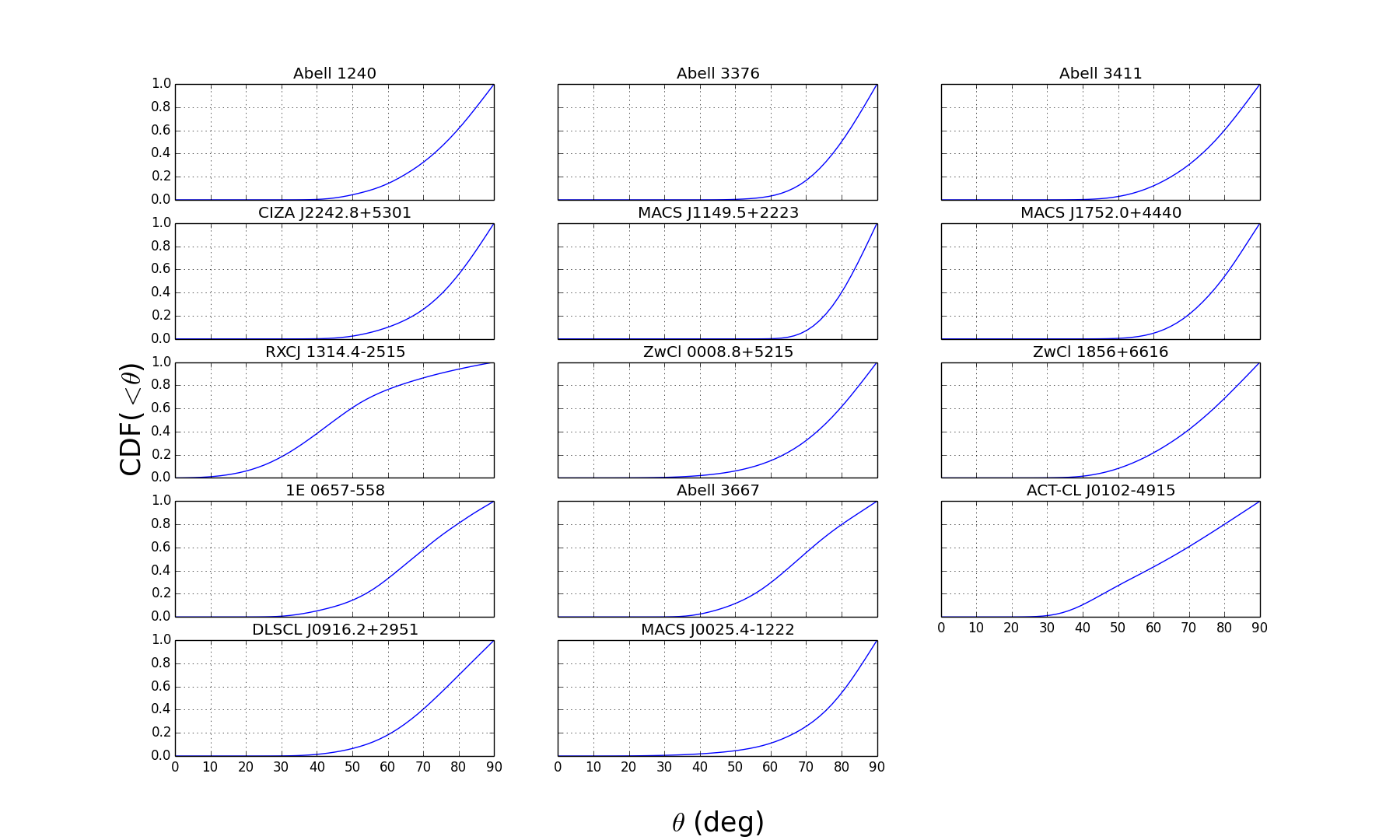}}
\caption{CDF of $\theta$ for each cluster at different pericenters. Each panel uses the PDF from the corresponding panel of Figure~\ref{fig-thetaconstraints} with the exception of Abell 3667 and El Gordo which use their respective curves in Figure~\ref{fig-revised_inputs}.}
\label{fig-CDFplots}\end{figure*}

A typical use of this method would be to find confidence intervals on the viewing angle. To facilitate this, we integrate each of the likelihoods plotted in  Figure~\ref{fig-thetaconstraints} to produce a cumulative distribution function (CDF) for each cluster; these are plotted in Figure~\ref{fig-CDFplots}. Table~\ref{tab-results} lists the 68\% and 95\% confidence intervals.

\begin{table}
\centering
\begin{tabular}{ccc}
System & 68\% CI & 95\% CI\\
\hline

 & MCC Clusters & \\
\hline
Abell 1240 & $\theta>70^\circ$ & $\theta>51^\circ$\\
Abell 3376 & $\theta>75^\circ$ & $\theta>62^\circ$\\
Abell 3411 & $\theta>71^\circ$& $\theta>53^\circ$\\
Ciza J2242 & $\theta>73^\circ$ & $\theta>54^\circ$\\
MACS J1149 & $\theta>78^\circ$ & $\theta>69^\circ$\\
MACS J1752 & $\theta>74^\circ$ & $\theta>60^\circ$\\
RXCJ 1314 & $\theta>37^\circ$ & $\theta>19^\circ$\\
ZWCI 0008 & $\theta>70^\circ$ & $\theta>48^\circ$\\
ZWCI 1856 & $\theta>66^\circ$ & $\theta>46^\circ$\\
\hline
 & Other Clusters & \\
\hline
1E 0657 & $\theta>61^\circ$ & $\theta>40^\circ$\\
Abell 3667 & $\theta>63^\circ$& $\theta>47^\circ$\\
ACT-CL J0102 & $\theta>57^\circ$ & $\theta>36^\circ$\\
DLSCL J0916 & $\theta>67^\circ$ & $\theta>48^\circ$\\
MACS J0025 & $\theta>73^\circ$ & $\theta>51^\circ$\\
\end{tabular}
\caption{Confidence intervals on viewing angles based on the CDF from Figure~\ref{fig-CDFplots}}
\label{tab-results}
\end{table}

\subsection{Comparison with other constraints}\label{subsec-constraintcomparison}

The MCMAC code \citep{Dawson2012} implements the modern version of the timing argument described in Section~\ref{sec-intro}. Table 3 of \citet{Dawson2012} lists confidence intervals for his parameter $\alpha$, which is the complement of our $\theta$: $\theta = 90^\circ - \alpha$. Dawson's result for DLSCL J0916, in our parametrization, is that $77^\circ>\theta>12^\circ$ at 95\% confidence. Our analog method produces a substantially stronger preference for the plane of the sky, $\theta>67^\circ$ at 95\% confidence.  Applying MCMAC to additional clusters, we find that MCMAC never prefers the plane of the sky, while the analog method often does.  This is due to MCMAC's assumption that the relative velocity vector is parallel to the separation vector. This assumption makes it impossible for the separation vector to be in the plane of the sky for systems with nonzero observed line-of-sight velocity difference. However, merging systems in BigMDPL generally have relative velocity vectors oblique to the separation vector. This results in the plane-of-sky model not being strongly suppressed unless the observed relative velocity is many hundreds of km/s. Without this suppression, the plane-of-sky model is the most likely based on the geometry of randomly placed observers. An instructive counterpoint is provided by those clusters with substantial line-of-sight velocity difference. The Bullet cluster has 
615 km/s observed relative velocity, and in this case our analog method does find a suppressed likelihood for the plane-of-sky model, agreeing with the MCMAC result \citep{Dawson2012} that the separation vector is most likely $\approx20^\circ$ from the plane of the sky.\footnote{RXCJ 1314.4-2515 would presumably be an even better example, but there is no published MCMAC result for this cluster.} 

A number of papers have built on MCMAC by factoring in the polarization of the radio relic; according to \citet{Ensslin1998} high polarization favors a separation vector near the plane of the sky. \citet{GolovichZwCl0008}, for example, find that $\theta$ in ZwCl 0008.8+5215 is poorly constrained by vanilla MCMAC, but that polarization tightens the constraint and pushes it toward larger $\theta$: $59<\theta<83.4$ at 68\% confidence.  Still, because MCMAC excludes plane-of-sky models, the \citet{GolovichZwCl0008} likelihood cuts off sharply as $\theta\to 90^\circ$, while ours peaks there. A similar story plays out with MACS J1149.5+2223 \citep{GolovichMACS1149} and Abell 3376 \citep{Monteiro2017Abell3376}. If one were to use the analog method rather than MCMAC as a starting point for applying the polarization-based likelihood factor, plane-of-sky geometry would be substantially more strongly favored. For El Gordo the analog method agrees with the MCMAC plus polarization analysis \citep{Ng2015} in favoring a model with the separation vector slightly out of the plane of the sky; the difference is that the analog method disfavors the plane-of-sky model only slightly whereas MCMAC excludes it entirely.

\section{Summary and Discussion}\label{sec-discussion}
We used the BigMDPL cosmological n-body simulation to constrain the angle $\theta$ between the subcluster separation vector and the line of sight, as a function of observed masses, relative radial velocity, and projected separation. This method naturally marginalizes over a cosmologically motivated range of impact parameter and initial velocity, and naturally incorporates the physical effects of large-scale structure, substructure, and dynamical friction.
For most observed mergers, a separation vector in the plane of the sky is favored. This may not be surprising because geometry dictates that a randomly placed vector is most likely to be perpendicular to an observer. However, this result disagrees with the MCMAC code which implements a modern version of the timing argument: MCMAC analyses routinely exclude the plane-of-sky model even where {\it nearly} plane-of-sky models are favored. We trace this to the MCMAC assumption that the merger is head-on, as follows. Observed systems typically have slightly nonzero $\Delta v_r$, and if velocity vectors are assumed parallel to the separation vector, one can match the observations only by tilting the separation vector at least slightly away from the plane of the sky. The analog method erases this artifact by marginalizing over a realistic range of systems where the velocity vectors are oblique to the separation vector. The analog method restores the $\theta=90^\circ$ model to most likely status for most observed systems, and with a substantially tighter constraint than could be obtained by the randomly placed vector argument alone.  

We explored how these results depend on the measured values of $\Delta v_r$ and $d_{\rm proj}$. An instructive outlier is RXC J1314, which has the largest $\Delta v_r$ and the smallest $d_{\rm proj}$, implying a separation vector that points substantially away from the plane of the sky. We also presented constraints for systems over a coarse grid of ($\Delta v_r$,$d_{\rm proj}$) values, which could be used for systems yet to be discovered. We found that gridding over the subcluster masses is unnecessary for this purpose because the mass effect is small.

We found sufficient numbers of analogs for most observed systems, even the Bullet; the exceptions were Abell 3667, which we attribute to inflated dynamical mass estimates, and El Gordo, which has a high lensing mass for a redshift of 0.87. MCMAC, by comparison, is immune to these considerations. Very tight observational error bars could also potentially limit the number of analogs found. MCMAC works even with very narrow ranges on its input parameters because it considers a continuous family of models, whereas the analog method could overcome this only by moving to an n-body simulation with a larger volume. 

A limitation of the current paper is that we used the BigMDPL halo coordinates and projected them, whereas lensing observations first project the mass distribution and then find a centroid (or similar summary statistic). Ideally, simulation {\it particle} data would be projected to make a more direct analog to lensing constraints. However, this would require much more computation, and particle data are not often publicly available. Further work will be needed to determine whether adding this level of realism would have any impact on the constraints yielded by the analog method. Adding gas to the simulations could also increase the fidelity of the analog method.

As the analog method is developed further, more care could be taken to translating the observations into analog selection criteria; for example if the projected separation was measured using BCGs one could use a simulation with galaxies and actually select on BCG separation. However, the constraints we developed using observed masses, velocity, and projected separation alone are typically broad and may not change appreciably with further detailed refinement of the selection criteria. A more fruitful direction may be to factor in additional constraints, from polarization for example. Thus, the analog method could eventually replace the timing argument as the basic likelihood upon which additional likelihoods are layered. The analog method could also be extended to infer additional merger parameters such as time since pericenter, and it could be used to identify boxes for zoom simulations that focus on additional merger details at higher resolution. 

\acknowledgments

This work was supported in part by NSF grant 1518246. This paper grew out of an idea originally suggested by Will Dawson, and we thank him for many useful discussions. We also thank Nate Golovich and Craig Lage for useful discussions.
The CosmoSim database used in this paper is a service by the Leibniz-Institute for Astrophysics Potsdam (AIP).
The MultiDark database was developed in cooperation with the Spanish MultiDark Consolider Project CSD2009-00064.

\bibliography{ms}

\end{document}